\documentclass[a4paper,12pt]{article}

\usepackage{amsmath}
\usepackage{graphicx}
\usepackage{hyperref}
\usepackage{color}
\usepackage{subfigure}
\usepackage{cite}

\def\beq{\begin{equation}}
\def\eeq{\end{equation}}
\def\barr{\begin{array}}
\def\earr{\end{array}}
\def\dis{\displaystyle}
\def\ltap{\, \raisebox{-.4ex}{\rlap{$\sim$}} \raisebox{.4ex}{$<$} \,} 
\def\gtap{\, \raisebox{-.4ex}{\rlap{$\sim$}} \raisebox{.4ex}{$>$} \,} 
\def\gev{\, {\rm GeV}} 
\def\tev{\, {\rm TeV}} 

\begin{document}
\begin{center} 
\begin{Large}\textbf{Probing Top Anomalous Couplings at the Tevatron and the Large Hadron Collider}\end{Large}\\
\vspace*{20pt} 
Debajyoti Choudhury and Pratishruti Saha\\
\vspace*{5pt}
\begin{footnotesize}Department of Physics and Astrophysics, University of Delhi, Delhi 110007, India.\end{footnotesize}\\
\end{center} 

\vspace*{40pt}

\begin{abstract} 
\noindent
Chromomagnetic and chromoelectric dipole interactions of the top quark
are studied in a model independent framework. Limits are set on the
scale of new physics that might lead to such contributions using
latest Tevatron measurements of the $t\bar t$ cross-section. It is 
demonstrated that the invariant mass distribution is a sensitive probe. 
Prospects at the LHC are examined. It is  shown that, for unitarized 
amplitudes, an increase in the LHC energy is of little importance, 
while the accumulation of luminosity plays a crucial role.
\newline

\vspace*{25pt}
\noindent
\texttt{PACS Nos:14.65.Ha,14.70.Dj,12.38.-t} \\ 
\texttt{Key Words:top,anomalous,Tevatron,LHC, etc.} 
\end{abstract}

\vspace*{40pt}

%%%%%%%% INTRODUCTION %%%%%%%%%%%%%%%%%%%%%%%%%%%%%%%%%%%%%%%%%%%%%%%%%%%%%%%%%%
\section{Introduction}

The Standard Model (SM) embodies our current
understanding of the fundamental constituents of the universe and
their interactions. This model is well tested up to the energy scale of
a few hundred GeVs and experiments have shown many of its predictions
to be astoundingly accurate. However, in spite of this stupendous
success, certain questions remain unanswered. Among them are questions
regarding the mechanism responsible for giving masses to fundamental
particles.

Within the SM, the generation of masses 
is explained by the spontaneous breaking of the electroweak
symmetry and the Higgs mechanism. However, no such Higgs 
scalar has been found yet. Furthermore, the SM 
fails to explain why, even though the underlying mechanism is the
same, there is a difference of six orders of magnitude between the masses
of the lightest fermion (electron) \footnote{We do not consider the 
neutrinos here as even the nature of their mass term is, as yet, 
uncertain. Were they to be purely Dirac ones, the hierarchy worsens.} 
and the heaviest one (top quark).  The Yukawa
couplings of the fermions with the Higgs are parameters in the
SM and cannot be explained or predicted by the theory.

It is possible to bring about electroweak symmetry breaking (EWSB)
without introducing a new fundamental field such as the Higgs
\cite{DyEWSB}. What is important to note is that, any theory that
provides a mechanism for generation of masses must have a large
coupling to the top quark. Consequently, even in the absence of an
actual observation of the Higgs boson, experiments with the top may be
used to probe the EWSB mechanism. Not only is the top quark the
heaviest particle in the SM with a mass $\sim$175 GeV,
its mass differs widely from those of the other fermions (the next
heaviest is the $b$-quark with a mass of 4.2 GeV \cite{PDG}). This
prompts us to examine whether the top quark has couplings different
from and in addition to those of the other quarks.

At the Tevatron proton-antiproton collider at Fermilab, the
mass and charge of the top quark have been measured reasonably well
\cite{CDF-D0,Top_Charge}. But, the high threshold for top production has
meant that its couplings are still not well measured. The possibility
that the top quark has anomalous couplings is still open. Once the
Large Hadron Collider at CERN comes into operation, precise
measurements of top couplings and detection of anomalous couplings
will become possible.

Various anomalous couplings of the top have been discussed in
Ref.\cite{eff_terms}. Of these, the ones that pertain to the QCD-sector
would be expected to modify the production rates significantly and,
thus, to be probed during the early phase of the LHC. On the other
hand, modifications of the electroweak couplings would play only a
sub-dominant role in $t \bar t$ production and it is the decay
patterns that would be affected more by them. Consequently, a search for
the latter type would require both a thorough understanding of the detector
as well as the accumulation of large statistics. Given this,
we concentrate here on the former set.

Large anomalous couplings may arise in a plethora of models. 
Prominent among these are scenarios of dynamical EWSB. 
Mechanisms for dynamic breaking of electroweak symmetry through the
formation of $t\bar t$ condensates are discussed in
Ref.\cite{tt_condensate}. Such scenarios require the the top quark to
have non-QCD `strong' interactions and give rise to interaction terms
such as $\bar tt \bar tt$ and $\bar tt\bar bb$ which then contribute to the
higher order corrections to the $ttg$ vertex.
Contributions may also arise from theories with additional heavy
fermions that couple to the top. Examples include, but are not limited
to, Little Higgs models~\cite{lh0,lh_ew} or models with extra
spacetime dimensions~\cite{acd_ued,ued_others,Barbieri_ed}. Another
possibility is the SM augmented by color-triplet or color-sextet
scalars that have Yukawa couplings with the top-quark~\cite{diquark}.

In a model independent framework, the lowest-dimensional anomalous coupling
of the top with the gluon can be parametrized by extra terms in the
interaction Lagrangian of the form 
\begin{equation}
{\cal L}_{int} \ni 
\frac{g_s}{\Lambda} \, F^{\mu\nu}_a  \; 
\bar t  \sigma_{\mu\nu} (\rho + i \, \rho' \, \gamma_5) \, T_a \, t 
\label{lagrangian}
\end{equation}
where $\Lambda$ denotes the scale of the effective theory.  While
$\rho$ represents the anomalous chromomagnetic dipole moment of the
top, $\rho'$ indicates the presence of a ($CP$-violating)
chromoelectric dipole moment. Within the SM, $\rho'$ is non-zero only
at the three-loop level and is, thus, tiny. $\rho$, on the other hand,
receives a contribution at the one-loop level and is ${\cal
  O}(\alpha_s/\pi)$ for $\Lambda \sim m_t$. The evidence for a larger
$\rho$ or $\rho'$ would thus be a strong indicator of new physics
lurking nearby. Whereas both $\rho$ and $\rho'$ can, in general, be
complex, note that any imaginary part thereof denotes absorptive
contributions and would render the Lagrangian non-Hermitian. We desist
from considering such a possibility.

The phenomenological consequences of such anomalous couplings have been 
considered earlier in Ref. \cite{previous}. However, we reopen the issue
in light of the improved measurements of top quark mass and $t\bar t$
cross-section and the first reported measurement of $t\bar t$ invariant mass.

%%%%%%%%%%%%%%%%%%%%%%%%%%%%%%%%%%%%%%%%%%%%%%%%%%%%%%%%%%%%%%%%%%%%%%%%%%%%%%%%

%%%%%%%% ANALYTIC %%%%%%%%%%%%%%%%%%%%%%%%%%%%%%%%%%%%%%%%%%%%%%%%%%%%%%%%%%%%%%
\section{Analytic Calculation}

The inclusion of a chromomagnetic moment term leads to a modification
of the vertex factor for the usual $ttg$ interaction to
$ig_s[\gamma^{\alpha} +
 (2 \, i \, \rho / \Lambda) \, \sigma^{\alpha\mu}k_{\mu}]T^a$ where 
$k$ is the momentum of the gluon coming
into the vertex.  An additional quartic interaction involving two top
quarks and two gluons is also generated \footnote{Note that, while this
vertex has occasionally been dropped or modified in literature, its inclusion 
is necessary for the $g g \to t \bar t$ amplitude to be a gauge 
invariant one.} 
with the corresponding vertex factor being
$(2 \, i \,g_s^2 \, \rho / \Lambda) \, f_{abc}\sigma^{\alpha\beta}T^c$. 
The changes in the presence of the 
chromoelectric dipole moment term are analogous, 
with $\rho$ above being replaced by $ (i \, \rho' \, \gamma_5)$.

At a hadron collider, the leading order 
contributions to $t\bar t$ production come from
the $q\bar q \rightarrow t\bar t$ and $gg \rightarrow t\bar t$
sub-processes. Summing (averaging) over spin and color degrees 
of freedom and defining $\Theta_{\pm}$ = $1
\pm \beta^2 \cos^2\theta$ where, $\beta = \sqrt{1-4m_t^2/\hat s}$ and
$\theta$ are, respectively, the velocity and scattering angle of the
top in the parton center-of-mass frame, the differential cross-sections
can be expressed as 
\beq
\barr{rcl}
\dis
\left(\dfrac{2\hat s}{\pi\alpha_s^2\beta}\right)\dfrac{d\hat\sigma_{q\bar q}}{d\cos\theta} & = & \dis
\dfrac{2}{9}\Theta_+ + \dfrac{8}{9}\dfrac{m_t^2}{\hat s} 
\\[2ex]
& + & 
\dfrac{32 \, \rho \, m_t}{9 \, \Lambda} + 
\dfrac{8 \, \rho^2}{9 \, \Lambda^2} \, 
\left( \hat s \, \Theta_- + 4 \, m_t^2\right) + 
\dfrac{8 \, \rho'^2}{9 \, \Lambda^2} \, 
\left( \hat s \, \Theta_- - 4 \, m_t^2\right)\, ,

\\[3ex]
\dis 
\left(\dfrac{2\hat s}{\pi\alpha_s^2\beta}\right)\dfrac{d\hat\sigma_{gg}}{d\cos\theta} & = & \dis
\dfrac{2}{3\Theta_-} \left(1 + \dfrac{4m_t^2}{\hat s} + \dfrac{m_t^4}{\hat s^2}\right)
-
\left(\dfrac{1}{3} + \dfrac{3}{16}\Theta_+ 
      + \dfrac{3 \, m_t^2}{2 \, \hat s} 
+ \dfrac{16 \, m_t^4}{3 \, \hat s^2} \, \dfrac{\Theta_+}{\Theta_-^2}
\right)
\\[2ex]
& +& \dis
\frac{\rho \, m_t}{\Lambda}
\left(-3 + \dfrac{16}{3\Theta_-}\right)
+ \frac{\rho^2}{\Lambda^2} \Bigg[\dfrac{7}{3}\hat s + m_t^2\left\{-6 + \dfrac{34}{3\Theta_-}\right\}\Bigg]
\\[2ex]
& + & \dis
\frac{\rho'^2}{\Lambda^2} \Bigg[\dfrac{7}{3}\hat s + \dfrac{2m_t^2}{3\Theta_-}\Bigg]
\\[2ex]
& + & \dis
\frac{\rho}{\Lambda}\,\left(\frac{\rho^2}{\Lambda^2} + \frac{\rho'^2}{\Lambda^2}\right) \, m_t \left(\dfrac{28}{3}\hat s - \dfrac{20}{3\Theta_-}m_t^2\right)
\\[2ex]
& + & \dis
\frac{4}{3}\left(\frac{\rho^2}{\Lambda^2} + \frac{\rho'^2}{\Lambda^2}\right)^2\,
   \left( \hat s^2\Theta_- - m_t^2\hat s + \dfrac{4}{\Theta_-}m_t^4\right)\, .
\earr
   \label{mesq}
\eeq
In each case, the first line refers to the SM result (we do not
exhibit the electroweak contribution for the $q \bar q$-initiated
process as it remains unaltered) and the rest encapsulate the
consequences of the anomalous dipole moments.

It should be noted that there are neither terms with an odd power of
$\rho'$ nor do the expressions show any possibility of a
forward-backward asymmetry, even in the presence of a non-zero
$\rho'$. The two facts are inter-related. 
With $\rho$ ($\rho'$) representing a parity-even (odd) operator, 
clearly, terms odd in $\cos\theta$ have to be proportional to 
odd powers of $\rho'$.
On the other hand, with the chromoelectric dipole moment being a 
$CP$-violating one, an odd power of $\rho'$ would denote a $CP$-odd
(and $T$-odd) observable. It is easy to see that no such observable can be
constructed out of the four momenta (the two initial state 
hadrons and the tops) alone.
Had we the ability to measure the polarizations, that possibility would
open up too \cite{T_odd_corr}. Although preliminary efforts 
in this direction are underway \cite{pol_study}, the accumulated
statistics is unlikely to be enough to look for this subtle effect. 
Similarly, the presence of absorptive parts of $\rho$ (or $\rho'$) 
would have allowed for the existence of such terms 
(essentially by changing the properties of the operator under
time-reversal).

A further feature is the growth of the cross-sections with energy as 
is expected in a theory with dimension-five (or higher) 
operators. While 
this may seem unacceptable on account of a potential loss of 
unitarity\footnote{Note that $\rho$ and $\rho'$ lead 
to identical unitarity-breaking terms, a consequence of the fact 
that the difference between them necessarily has to be 
translated to subdominant terms.}, 
one should realize that the theory of eqn.(\ref{lagrangian}) is only an
effective one and is expected to be superseded beyond the scale 
$\Lambda$. While unitarity may be restored by promoting 
$\rho$ ($\rho'$) from constants to form-factors with appropriate 
powers of $(1 + \hat s / \Lambda^2)$, this is an {\em ad-hoc} measure 
as the mechanism of unitarity restoration is intricately related to 
the precise nature of the ultraviolet completion. We desist from doing this 
with the {\em a posteriori} justification that the limits of sensitivity 
for $\rho$ (as described in the next section) are far beyond the typical 
subprocess energies ($\hat s/\Lambda \ll$ 1). Furthermore, note that the 
terms of ${\cal O}(\rho)$ do respect partial wave unitarity. 
These terms \footnote{And, similarly, those of 
${\cal O}(\rho^3, \rho \rho'^2)$.} 
appear only as a result of interference between pure QCD and the 
dipole contributions,
and owing to the different chirality structures of the operators, have to be 
proportional to $m_t$ 
(thereby suppressing the growth with energy).

%%%%%%%%%%%%%%%%%%%%%%%%%%%%%%%%%%%%%%%%%%%%%%%%%%%%%%%%%%%%%%%%%%%%%%%%%%%%%%%%

%%%%%%%% NUMERICAL %%%%%%%%%%%%%%%%%%%%%%%%%%%%%%%%%%%%%%%%%%%%%%%%%%%%%%%%%%%%%
\section{Numerical Analysis}

Armed with the results of the previous section, we compute the
expected $t\bar t$ cross-section at the Tevatron as well as the
LHC. To this end, we use the CTEQ6L1 parton distribution
sets~\cite{CTEQ} with $m_{t}$ as the scale for both factorization as
well as renormalization. For a consistent comparision with the 
cross-section measurement reported by the CDF collaboration\cite{CDF_top_csec},
we use $m_{t} = 172.5$ GeV for the Tevatron analysis. For the LHC analysis, 
though, we use the updated value of $m_{t} = 173.1$ GeV, obtained as a result
of the combined CDF+D\O{} analysis\cite{CDF-D0}. To incorporate the
higher order corrections absent in our leading order results, we use
the $K$-factors at the NLO+NLL level \footnote{In the absence of a
  similar calculation incorporating anomalous dipole moments, we use
  the same $K$-factor as obtained for the SM case. While this is not
  entirely accurate, given the fact that the color structure is
  similar and drawing from experience with analogous calculations for
  higher dimensional operators~\cite{NLO_highdim}, the error associated with
  this approximation is not expected to be large.} calculated by Cacciari
et. al. \cite{Cacciari}.  Once this is done, the theoretical errors in
the calculation owing to the choice of PDFs and scale are approximately
7-8\% for the Tevatron and 9-10\% for the LHC \cite{Cacciari}.
For the LHC operating at 7 TeV, though, we use the approximate NNLO cross
section as reported in Ref.~\cite{Kidonakis}.

\subsection{Tevatron Results}
\begin{figure}[!htbp]
\centering
\subfigure[]
{
\includegraphics[scale=0.88]{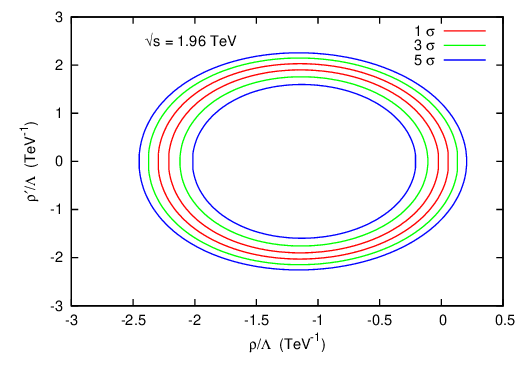}
}
\subfigure[]
{
\includegraphics[scale=0.88]{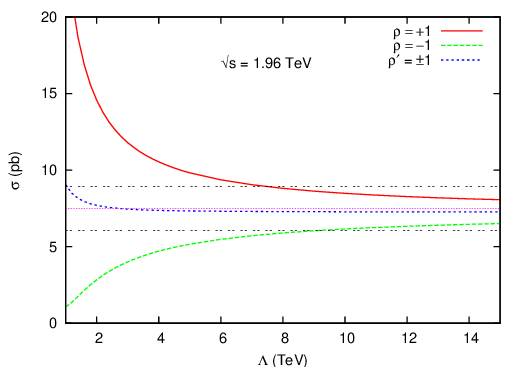}
}
\caption{\em {\em (a)} 
The region in ($\rho/\Lambda$)-($\rho'/\Lambda$) plane allowed
 by the Tevatron data \cite{CDF_top_csec} at the 1-$\sigma$, 3-$\sigma$
 and 5-$\sigma$ level. {\em (b)} 
  $t\bar t$ production rates for the Tevatron ($\sqrt{s}$ =
  1.96 TeV). The horizontal lines denote the CDF central value and the
  3-$\sigma$ interval \cite{CDF_top_csec}. }
\label{fig:contour}
\end{figure}

At the Tevatron, the dominant contribution accrues from the $q \bar q$
initial states, even on the inclusion of the dipole moments.  While
the ${\cal O}(\rho^2, \rho'^2)$ terms in $d \sigma / d \cos\theta$ are
always positive (see eqn.\ref{mesq}), the flat ${\cal O}(\rho)$ term
can flip sign with $\rho$. This implies that for $\rho > 0$, the
change in the cross-section, $\delta \sigma$, is positive. This severely
constrains any deviation of the anomalous couplings in that
direction. On the other hand, for $\rho < 0$, large cancellations may
occur between various pieces of the cross-section. Consequently,
substantial negative $\rho$ could be admitted, albeit correlated with a
substantial $\rho'$. This is exhibited by Fig.\mbox{\ref{fig:contour} ($a$)} which
displays the parameter space that is still allowed by the Tevatron
data, namely\cite{CDF_top_csec}
\beq
  \sigma_{t\bar t}(m_t = 172.5 \gev) = (7.50 \pm 0.48) \, {\rm pb}\,.
\eeq
The near elliptical shape of the contours is but a reflection 
of the fact that, at the Tevatron, the $q \bar q$ contribution far 
supersedes the $gg$ one. 

Having seen the extent to which cancellations may, in principle, be
responsible for hiding the presence of substantial dipole moments, we
now restrict ourselves to the case where only one of $\rho$ and
$\rho'$ may be non-zero. While this might seem a gross simplification,
it is not really so. For one, with the chromomagnetic moment
manifesting itself at ${\cal O}(\Lambda^{-1})$ and the chromoelectric
moment appearing in the cross-sections only at ${\cal O}(\Lambda^{-2})$,
it is obvious that, for large $\Lambda$, the
former would, typically, leave a larger imprint. Secondly, it is
extremely unlikely that the couplings conspire to be just so that
large cancellations take place.  This is particularly true because, for
a generic underlying ultraviolet completion, one would expect the
chromoelectric moment operator to appear at a higher order of perturbation
than the chromomagnetic one. On the other hand, the situation could be
reversed if there is an underlying symmetry (\`{a} la the
symmetry proposed in Ref.\cite{neutrino_magmom} to account for neutrino magnetic
moments) that prevents $\rho$ from appearing while allowing a non-zero
$\rho'$.

If only one of the two couplings are to be non-zero, we may rescale
$\rho, \rho' = 0, \pm 1$ and, thus, reduce the parameter space to one
dimension ($\Lambda$). Of course, $\rho' = \pm 1$ are equivalent.
Fig.\mbox{\ref{fig:contour} ($b$)} exhibits the corresponding dependence of
the total cross-section at the Tevatron on $\Lambda$ for various
combinations of $(\rho, \rho')$. For $\rho = +1,-1$, the near-monotonic
dependence on $\Lambda$ is reflective of the dominance of the ${\cal
  O}(\rho/\Lambda)$ term. This is particularly true for $\Lambda/\rho
\gtap 3 \tev$. 

The low sensitivity to the chromoelectric moment is
understandable in view of the fact that the corresponding contribution
is suppressed by at least $\Lambda^2$. Furthermore, unlike in the case
of the chromomagnetic moment, the $q \bar q \to t \bar t$ cross-section
in this case suffers an additional cancellation owing to the
chirality structure (see eqn.\ref{mesq}). With $\hat s$ at the
Tevatron being only slightly greater than $4 \, m_t^2$, this
cancellation is quite significant.

\begin{figure}[!htbp]
\centering
\includegraphics[scale=0.90]{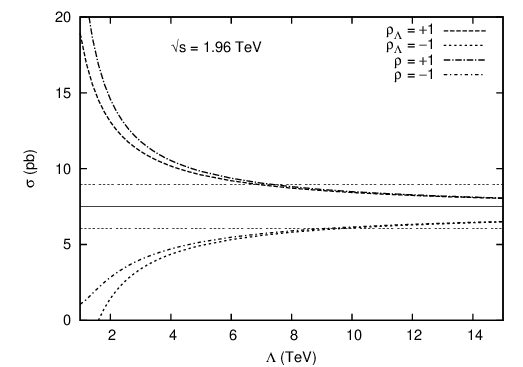}
\caption{\em Comparison of production rates obtained at the Tevatron with truncated cross-sections (up to ${\cal O}(\Lambda^{-1})$; denoted by subscript $\Lambda$ in the key) and full cross-sections(all orders in $\Lambda$).}
\label{fig:comp_trunc}
\end{figure}

Note that, for $\rho \neq 0$, while Fig.\mbox{\ref{fig:contour} ($a$)}
shows a second range (close to $\rho / \Lambda \sim -2.2 \tev^{-1}$)
consistent with experimental observation, the existence of the same is 
not apparent in Fig.\mbox{\ref{fig:contour} ($b$)}. As can be easily 
appreciated, for smaller $\Lambda$, the ${\cal O}(\rho^2/\Lambda^2)$ term
gets progressively more important, leading to a rapid growth in the 
total cross-section for $\rho > 0$ and a cancellation between the two
leading $\Lambda$ dependent terms for $\rho < 0$.
Consequently, for smaller values of $\Lambda$ (not shown on the plot),
the lowest curve in Fig.\mbox{\ref{fig:contour} ($b$)} would
actually suffer a turnaround, rendering it consistent with the measurements
for a certain range of $\Lambda$.
However, one should not be lead on too far by this. It is the 
${\cal O}(\Lambda^{-2})$ contributions that are largely responsible
for this second region of consistency. On the other hand, 
the Lagrangian considered in eqn.\ref{lagrangian} contains only the 
lowest dimensional anomalous operators of an effective
theory. Higher dimensional operators \cite{dim_6}, if included in the 
Lagrangian, could change the behaviour of the cross-sections and hence the
conclusions drawn from Fig.\mbox{\ref{fig:contour} ($b$)}. A closer examination of
this issue (see Fig.\ref{fig:comp_trunc}) reveals that that were we to neglect
${\cal O}(\Lambda^{-2})$ terms in eqn.\ref{mesq}, the shape of the curves would
indeed change considerably
\footnote{While it may be argued that, in principle, 
  some as yet   unknown symmetry could render such higher order terms in
  eqn.\ref{lagrangian} to be very small, we feel that such an
  eventuality would be a very artificial one.}, 
but the limits on $\Lambda$ for either of $\rho = \pm 1$ would hardly alter.
In other words, the sensitivity limits are overwhelmingly dominated by the
${\cal O}(\rho/ \Lambda)$ terms, thus making them quite robust. In fact,
the change in the limits from inclusion of higher-order terms are well below
the theoretical errors from sources such as the dependence on the
factorization/renormalization scales, choice of PDF etc.  To be
quantitative, $\Lambda \ltap 7400 \gev$ can be ruled out at 99\%
confidence level for the $\rho = +1$ case. For $\rho = -1$ on the
other hand, $\Lambda \ltap 9000 \gev$ can be ruled out at the same
confidence level. One expects similar sensitivity for $\rho=+1$ and $\rho=-1$.
The difference essentially owes its origin to the
slight discrepancy between the SM expectations (as computed with our
choices) and the experimental central value.
Of course, restricting to ${\cal O}(\Lambda^{-1})$ eliminates $\rho'$ altogether.
However, sensitivity to $\rho'$ may still be obtained by including absorptive
pieces and/or by considering polarised scattering.

\subsection{LHC Sensitivity}
\begin{figure}[!htbp]
\centering
\subfigure[]
{
\includegraphics[width=2.0in,height=2.5in]{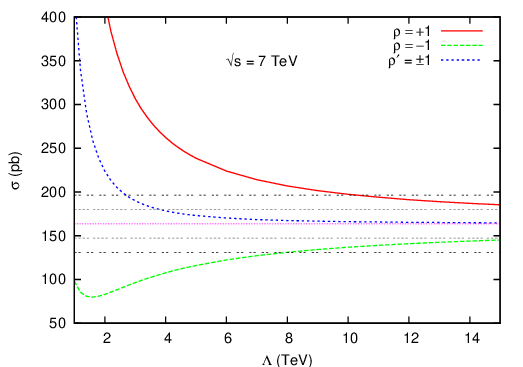}
}
\subfigure[]
{
\includegraphics[width=2.0in,height=2.5in]{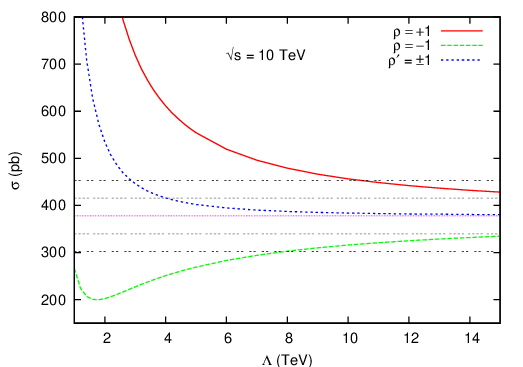}
}
\subfigure[]
{
\includegraphics[width=2.0in,height=2.5in]{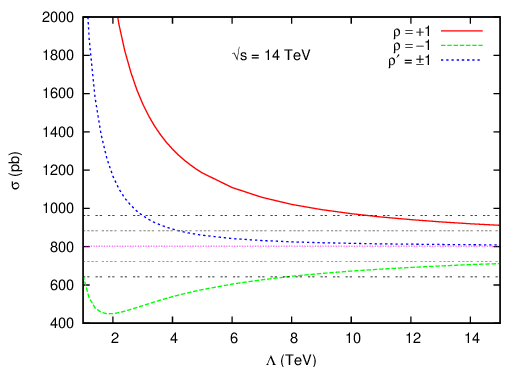}
}
\caption{\em $t\bar t$ production rates for the LHC as a function of
  the new physics scale $\Lambda$. Panels from left to right
  correspond to $\sqrt{s} = 7, \, 10, \, 14$ TeV. 
  The horizontal lines show the SM expectation and the 10\% and
  20\% intervals as estimates of errors in the measurement \cite{CMS-PAS}.}
\label{fig:tot_LHC}
\end{figure}

At the LHC, it is the $gg$ flux that rules the roost, especially at 
smaller $\hat s$ values.  Moreover, 
at high center-of-mass energies, 
the gluon-initiated cross-sections grow as $\hat s/\Lambda^4$, whereas 
the $q\bar q$-initiated cross-sections remain, at best,  constant with $\hat s$.
Consequently, it is fair to say that the $gg \to t \bar t$ subprocess 
dominates throughout. In Fig.\ref{fig:tot_LHC}, we present the 
corresponding cross-sections
at the LHC as a function of $\Lambda$ for various values of 
the proton-proton center-of-mass energy $\sqrt{s}$. 
In the absence of any data, we can only compare these with the 
SM expectations and the estimated errors. Experimental errors 
due to systematic and statistical
uncertainties are expected to be between 20 and 30
percent for an integrated luminosity of 20 ${\rm pb}^{-1}$ at
$\sqrt{s}$ = 10 TeV \cite{CMS-PAS} (the errors for other values of 
$\sqrt{s}$ are similar) and dominate the theoretical errors 
quoted earlier. Since experimental errors are
expected to decrease with better calibration of the detectors and
increase in statistics, we choose to display 10\% (optimistic) 
and 20\% error bars for comparison. 

Setting bounds on the chromoelectric dipole moment now becomes
possible. Much of this is due to the fact that the $gg \to t \bar t$
amplitude is not as chirally suppressed as the $q \bar q \to t \bar t$
one (see eqn.\ref{mesq}). For non-zero $\rho'$, even an early run
of the LHC with $\sqrt{s}$ = 7 TeV (Fig.\mbox{\ref{fig:tot_LHC} $a$}) 
would be sensitive to $\Lambda \ltap 2700 \gev$. Unfortunately, 
the improvement of the sensitivity with the machine energy is marginal
at best. 

As for the chromomagnetic moment, the story is more complicated.
For $\rho = +1$, naively a sensitivity up to about
$\Lambda \sim 10 \tev$ could be expected.
A higher operative energy for the LHC renders it more suitable to the
chromomagnetic moment as long as $\rho = +1$. This growth in sensitivity 
is a reflection of the growing importance of the higher order 
(in $\rho/\Lambda$) terms as the energy is increased 
\footnote{These pieces in the cross-section do not fall off with $\hat s$}.
However, the magnitude of the increase in sensitivity with the $pp$ 
center-of-mass energy is small. This can be understood by looking at 
eqn.\ref{mesq}. The SM as well as the ${\cal O}(\rho/\Lambda)$ terms
in the cross-section fall with $\hat s$. At lower values of $\hat s$,
the SM piece falls faster. However, in the higher $\hat s$ regime, 
both have a similar behaviour and thus the increase in sensitivity 
that can be obtained by increasing the center-of-mass energy is only marginal.
 
For $\rho = -1$, on the other hand, the situation is reversed.
The contrasting behaviour is easy to understand in terms
of the constructive (destructive) interferences with the SM amplitude
in the two cases. The aforementioned cancellation between
various orders reappears in a more complicated guise even for the
$gg \to t \bar t$ case. On account of this, it appears
that the best that the LHC can do is to rule out (for $\rho = -1$)
$\Lambda \ltap 8 \tev$. This, however, should be compared with the Tevatron
results which have already ruled out $\Lambda \ltap 9 \tev$.
 
While the discussion above was based on the full cross-sections as
listed in eqn.\ref{mesq}, the situation changes somewhat if one
were to truncate contributions beyond ${\cal O}(\Lambda^{-1})$. In
Table.\ref{tab:lhc_lim}, we display the bounds on $\Lambda$ that may
be reached, with and without such a truncation, for the three
different stages of LHC operation. In reaching these bounds, we have
assumed that a 20\% deviation constitutes a discernible shift. 
For $\rho = +1$, a small increase in sensitivity is obtained by increasing
the machine energy, as with the untruncated cross-sections. The most 
interesting feature, however, is that for $\rho = -1$, with truncated 
cross-sections, one gets an improvement in the sensitivity on increasing 
in machine energy. This is contrary to the trend observed when the full 
cross-sections are considered. This is but yet another indication
of the importance of the terms of order $1/\Lambda^{2}$ and greater and their role in 
cancellations between various pieces in the cross-section at higher values of $\hat s$.
However, owing to the nature of the $gg$ and $q \bar q$--fluxes, most of 
the cross-section accrues from relatively smaller values of $\hat s$. Hence the
magnitude of the change is tiny.  

In fact, even with a several fold increase in the LHC energy, the sensitivity
would not increase by much. It is amusing to note that, were LHC a $p \bar p$ 
collider instead, the use of the full matrix-element-squared 
would have entailed a
substantial increase in the sensitivity with energy, although the situation 
for the
truncated case would have remained quite similar.
The use of the full matrix elements would have entailed using 
sub-process cross sections that grow with $\hat s$.
In the present situation, this has been offset by the
rapidly falling antiquark and gluon densities at large $x$-values.
For a $p \bar p$ collider, the $\bar q$ densities would not fall
off so fast, resulting in a growth of the cross section.
On the other hand, by truncating the cross section to ${\cal O}(\rho)$,
we essentially ensure that the sub-process cross-sections do 
not violate unitarity. The truncated sub-process cross sections
actually fall with $\hat s$ at approximately the same rate as the SM ones.
This ensures that the relative deviation does not grow even with
an increase in the $p \bar p$ center-of-mass energy.

\begin{table}
\begin{center}
\begin{tabular}{||c ||r|r ||r|r ||r|r||}
\hline
& \multicolumn{2}{|c||}{$\sqrt{s} = 7 \tev$}
& \multicolumn{2}{|c||}{$\sqrt{s} = 10 \tev$}
& \multicolumn{2}{|c||}{$\sqrt{s} = 14 \tev$}
\\[1ex]
\cline{2-7}
  & \multicolumn{1}{|c|}{Full} & \multicolumn{1}{|c||}{Trunc.}
  & \multicolumn{1}{|c|}{Full} & \multicolumn{1}{|c||}{Trunc.}
  & \multicolumn{1}{|c|}{Full} & \multicolumn{1}{|c||}{Trunc.}
\\
\hline
$\rho = +1$ & 10.20 & 9.20 & 10.45 & 9.25 & 10.50 & 9.30
\\[1ex]
\hline
$\rho = -1$ &  8.00 & 9.20 &  7.95 & 9.25 &  7.85 & 9.30
\\[1ex]
\hline
\end{tabular}
\end{center}
\caption{\em The values of $\Lambda$ (in TeVs) that would
 lead to a 20\% deviation 
in the cross-section from the SM expectations, for $\rho = \pm 1$ and 
$\rho' = 0$. In each case, the limits are shown as calculated with the 
full cross-sections of eqn.\ref{mesq} as well as those obtained from 
the expressions truncated at the ${\cal O}(\Lambda^{-1})$ level. 
}
\label{tab:lhc_lim}
\end{table}

\subsection{Use of Differential Distributions}
So far, we have only considered the total $t\bar t$ cross-section and
the deviations therein as a possible signal for the existence of
anomalous dipole moments. There exist other observables that can be
constructed and studied even in the complex detector environment of a
hadron collider.  The invariant mass distribution of the final state
particles is one such.  In 2009, the CDF collaboration reported the
first measurement of the $t\bar t$ invariant mass ($m_{t\bar t}$)
distribution \cite{CDF_mtt}.  This data can be used to put further
constraints on values of $\rho$ and $\rho'$.

\begin{figure}[!htbp]
\centering
\includegraphics[scale=1.0]{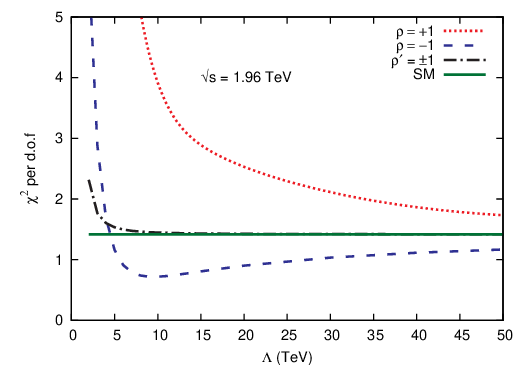}
\caption{\em $\chi^2$ per degree of freedom.}
\label{fig:chisq}
\end{figure}

The CDF collaboration reports the measurement in 9 bins between 0 and
1400 GeV (Fig.1 and Table III in Ref.\cite{CDF_mtt}) assumimg $m_t =
175$ GeV with 2.7 $fb^{-1}$ worth of data.  Note that the first bin
which extends in the range 0-350 GeV also has a non-zero number of
events, an artefact of experimental errors associated with the
reconstruction of the $t\bar t$ events as well as of effects due to
final state radiation. For our analysis, we exclude this bin. The
experimental effects are simulated so that the $m_{t\bar t}$
distribution for the SM matches with the CDF expectations.  As a
statistic, we consider a $\chi^2$ defined through

\[
   \chi^2 = \sum_{i = 2}^9 \left(\frac{\sigma^{\rm th}_i - \sigma^{\rm obs}_i}{\delta \sigma_i}
                          \right)^2
\]
where the sum runs over the bins and $\sigma^{\rm th}_i$ is the number 
of events expected in a given theory (defined by the values of 
$\rho, \rho', \Lambda$)
in a particular bin. $\sigma^{\rm obs}_i$ and $\delta \sigma_i$, on the other hand, 
are the observed event numbers and the errors therein. The
$\chi^2$ values thus obtained are plotted in Fig.\ref{fig:chisq} as function
of $\Lambda$. 

It is interesting to note that the $\rho = -1$ case gives a
better fit than the SM, over a
large range of $\Lambda$ values. Thus the data could be claimed to 
favour such a scenario! On the other hand, $\rho =  +1$ is now strongly 
disfavoured for much higher values of $\Lambda$, 
thereby exhibiting the aforeclaimed enhanced sensitivity 
of the $m_{tt}$ distribution. Even for the chromoelectric moment case 
($\rho' \neq 0$), the increase in sensitivity is evident. 
However, in all of this, we wish to
tread with caution.  This distribution has been constructed on the
basis of only 2.7 $fb^{-1}$ of data.  Robust limits may be obtained
once more statistics has been accumulated and a more realistic
simulation, with the inclusion of the effects of dipole moment
terms, has been carried out.

\begin{figure}[!htbp]
\centering
\subfigure[]
{
\includegraphics[width=7.9cm,height=7cm]{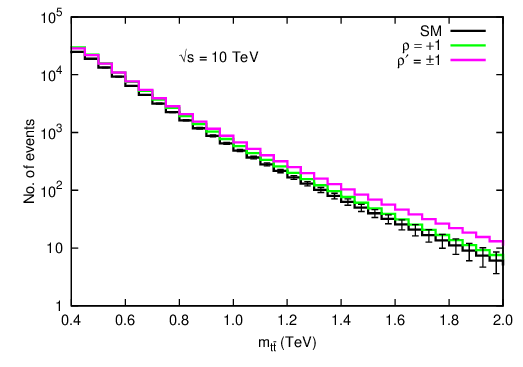}
}
\subfigure[]
{
\includegraphics[width=7.9cm,height=7cm]{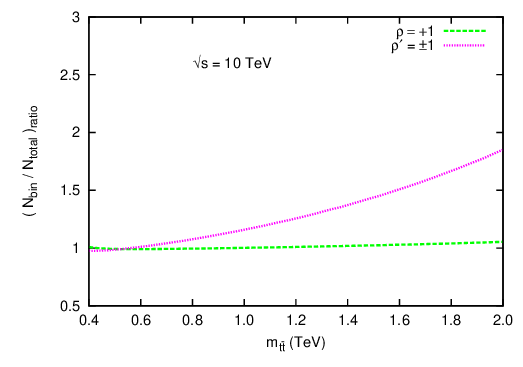}
}
\caption{\em {\em (a)} 
The $m_{t\bar t}$ spectrum for the LHC at $\sqrt{s}$ = 10 TeV along with the 
1-$\sigma$ Gaussian error bar for the SM.
{\em (b)} The ratio of the normalized $m_{t\bar t}$ spectra 
(see eqn.\ref{fig:norm_mtt}). 
In each case, the two anomalous sets refer to 
$(\rho, \rho', \Lambda) = (+1, 0, 11 \tev)$ and $(0, \pm 1, 3 \tev)$
respectively. An integrated luminosity of 300 $pb^{-1}$ has been assumed.}
\label{fig:mtt}
\end{figure}

At the LHC too, differential distributions will have a role to play 
in enhancing the sensitivity to different kinds of new physics scenarios
and discriminating between them.
From Fig. \mbox{\ref{fig:tot_LHC} ($b$)} one can see that, in $pp$ collisions
at $\sqrt{s}$ = 10 TeV, various combinations $(\rho,\rho') = (1,0), (0, \pm1)$
may give rise to positive deviations of the order of, say 15--20\% in the
total cross-section, albeit for wildly different values of $\Lambda$.
How can one distinguish between them? To answer this question, we once again 
turn to the invariant mass distribution and consider the full set
of expressions of eqn.\ref{mesq}, rather than the truncated ones. As
Fig.\mbox{\ref{fig:mtt} ($a$)} shows, the distributions do indeed diverge
significantly for large $m_{t\bar t}$. The two anomalous cross-sections 
depicted are roughly equal and deviate by approximately 20\% from the SM 
one. One might argue though that small differences in the spectrum could
{\em (a)} rise from various effects within the SM and/or experimental
resolutions and {\em (b)} get washed away as a result of poor statistics.
The second objection is countered by the observation that significant 
deviations are associated with a sizable event rate, even for a moderate
value of the integrated luminosity. This deviation is emphasized further
if one considers the ratio
\beq
    \left( \frac{1}{\sigma} \; \frac{d \sigma}{d m_{t\bar t}} \right)\bigg/
   \left( \frac{1}{\sigma_{SM}} \; \frac{d \sigma_{SM}}{d m_{t\bar t}} \right) \ .
    \label{fig:norm_mtt}
\eeq
This observable has the benefit of using normalized quantities
so that some of the systematic errors such as those due to 
luminosity measurements or lack of precise knowledge of the parton densities 
are largely removed. 
As a perusal of Fig.\mbox{\ref{fig:mtt} ($b$)} shows, the qualitative differences
between the cases stand out starkly. It should be noted that the 
fact of the normalized distribution for the $\rho = +1$ case being very 
closely aligned with the SM one is not accidental, but just a 
consequence of the fact that it corresponds to a larger value of 
$\Lambda$ compared to the other case. Consequently, the new physics 
contribution is dominated by the ${\cal O}(\rho/\Lambda)$ piece, the 
$m_{t\bar t}$ behaviour for which is quite similar to that for the SM piece.
The other parameter space point corresponds to a much smaller value 
of $\Lambda$ and the ${\cal O}(\rho^2/\Lambda^2)$ piece, which 
now plays an important role, has a very different $m_{t \bar t}$ 
dependence. 

Such distinctions can also be made using other kinematic variables such as
transverse momentum and difference in the rapidities of $t$ and $\bar t$.
However, they do not prove to be any more sensitive than the $m_{t\bar t}$ 
distribution.

%%%%%%%%%%%%%%%%%%%%%%%%%%%%%%%%%%%%%%%%%%%%%%%%%%%%%%%%%%%%%%%%%%%%%%%%%%%%%%%%

%%%%%%%% SUMMARY %%%%%%%%%%%%%%%%%%%%%%%%%%%%%%%%%%%%%%%%%%%%%%%%%%%%%%%%%%%%%%%
\section{Summary}

In a large class of models, the top-quark may have a substantial
anomalous chromomagnetic and/or chromoelectric dipole moment.  The
presence of such moments can lead to sizable deviations in $t\bar t$
production cross-sections at hadron colliders. A comparison with the 
Tevatron data, thus, serves to impose significant constraints on the 
parameter space. While fortuitous cancellations between different 
contributions to the total cross section can serve to allow for 
sizable values of such couplings, the adoption of the more conservative
effective field theory language implies that the Tevatron measurement of 
$\sigma_{t \bar t}$ alone implies that the corresponding new physics scale 
should be larger than approximately $7 \tev$. 

The use of the invariant mass spectrum presents an intriguing prospect. 
On the one hand, this results in far more severe restriction on 
positive values of the chromomagnetic moment. On the other, for 
moderate negative values of the same, the resultant spectrum is found 
to approximate the CDF data to a better extent than the SM does. 
While it is premature to claim this to be a discovery of a non-zero 
chromomagnetic dipole moment for the top, it, nevertheless, points 
to the need of understanding the spectrum better as well as to 
perform a more detailed analysis.

At the LHC, understandably, it would be possible to probe smaller 
values of such couplings.  However, the improvement is not likely 
to be a qualitative one (new physics scale $\Lambda \sim 10 \tev$) 
unless the errors on the $t \bar t$ cross sections, both 
theoretical and experimental, can be reduced significantly. 
Interestingly, with the relative deviation in the total cross section 
(due to new physics) being only very weakly dependent on the
$\sqrt{s}$, an increase in the accumulated luminosity is 
likely to lead to better constraints than an increase in the 
operating energy of the machine. Furthermore, a larger sample size would 
allow a more detailed use of the differential distributions (such as 
the $m_{t\bar t}$ one), leading to an even more enhanced sensitivity. Indeed,
it seems possible that this could even be used to distinguish between 
different operators that result in identical deviations in the 
total cross section. 

{\em Note added:} As this paper was being finalised
Ref.\cite{Okhuma_new} appeared as a preprint. Our expressions are in
agreement with those in Ref.\cite{Okhuma_new}.  Bearing in mind that
we use different computational methods, have different choices of
parton densities and use different methods to account for NLO effects,
our projected cross-sections can also be said to be in reasonable
agreement.

While this paper was being reviewed, the ATLAS~\cite{ATLAS-top-2011} and
the CMS~\cite{CMS-top-2011} collaborations announced their measurements of
$t \bar t$ cross-sections at the LHC. 
Combining an analysis of $36 \, (3) \, {\rm pb}^{-1}$ data in the 
semileptonic (dilepton) decay modes, CMS quotes 
$\sigma_{t \bar t} = 
[158 \pm 10 ({\rm stat}) \pm 15 ({\rm syst})\pm 6 ({\rm lumi})] \, {\rm pb}$. 
Similarly, analysing $35 \, {\rm pb}^{-1}$ data in similar modes, 
the ATLAS collaboration has 
$\sigma_{t \bar t} = 
[180 \pm 9 ({\rm stat}) \pm 15 ({\rm syst})\pm 6 ({\rm lumi})] \, {\rm pb}$. 
With the approximate next-to-next-to-leading order (NNLO) cross section, 
as calculated in Ref.~\cite{Kidonakis} being 
\[
\sigma_{t \bar t} = 
        \barr{lclcl}
             163^{+7}_{-5} ({\rm scale}) \pm 9 ({\rm PDF}) \, {\rm pb} 
                   &= & 163^{+11}_{-10}  \, {\rm pb} \ ,
         \earr
\]
both the measurements are in accordance with the SM value, and consistent with 
each other. It is interesting to note that the combined $2 \sigma$ 
experimental uncertainty is already at the 15\% mark. Thus, the 
bounds as derived from this data would already be competitive 
with those obtained from the Tevatron. However, given the theoretical
and experimental uncertainities, we feel that it is still too premature 
to derive any such stringent and definitive limits.

%%%%%%%%%%%%%%%%%%%%%%%%%%%%%%%%%%%%%%%%%%%%%%%%%%%%%%%%%%%%%%%%%%%%%%%%%%%%%%%%

%%%%%%%% ACKNOWLEDGEMENT %%%%%%%%%%%%%%%%%%%%%%%%%%%%%%%%%%%%%%%%%%%%%%%%%%%%%%%
\begin{center}
\begin{small}\textbf{ACKNOWLEDGEMENT}\end{small}\end{center}
DC acknowledges support from the Department of Science and
Technology, India under project number SR/S2/RFHEP-05/2006.
PS would like to thank CSIR, India for assistance under 
JRF Grant 09/045(0736)/2008-EMR-I.

%%%%%%%%%%%%%%%%%%%%%%%%%%%%%%%%%%%%%%%%%%%%%%%%%%%%%%%%%%%%%%%%%%%%%%%%%%%%%%%%

%%%%%%%% REFERENCES %%%%%%%%%%%%%%%%%%%%%%%%%%%%%%%%%%%%%%%%%%%%%%%%%%%%%%%%%%%%

%%%%%%%%%%%%%%%%%%%%%%%%%%%%%%%%%%%%%%%%%%%%%%%%%%%%%%%%%%%%%%%%%%%%%%%%%%%%%%%%

\end{document}